\documentclass[a4,12pt]{article}
cm \textheight 24.5 true cm \topmargin=-1mm

\begin{document}
\thispagestyle{empty}\setcounter{page}{1}
 \vskip40pt
\centerline{\bf Relevance of Quantum Mechanics in Circuit
Implementation of}
\centerline{\bf Ion channels in Brain Dynamics}

\vskip10pt

\centerline{\footnotesize Indranil Mitra$^1$, Sisir Roy$^{2,3}$}
\vskip5pt \centerline{\footnotesize $^1$National Center of
Biological Sciences} \centerline{\footnotesize TIFR ,UAS GKVK Campus
Bangalore-65 India } \vskip5pt
\centerline{\footnotesize$^{2}$Physics and Applied Mathematics Unit,
Indian Statistical Institute,}
\centerline{\footnotesize 203
Barrackpore Trunk Road, Kolkata 700108, India}

\centerline{\footnotesize$^{3}$School of Computational Sciences,
George Mason University} \centerline{\footnotesize 4400 University
Drive Fairfax, Virginia 22030,USA}

\centerline{\footnotesize$^{1}$e-mail: indranil@ncbs.res.in }
\centerline{\footnotesize$^{2}$e-mail: sisir@isical.ac.in}
\centerline{\footnotesize$^{3}$e-mail: sroy@scs.gmu.edu }

\vskip10pt

\abstract

With an increasing amount of experimental evidence pouring in from
neurobiological investigations, it is quite appropriate to study
viable reductionist models which may explain some of the features of
brain activities. It is now quite well known that the Hodgkin-Huxley
(HH) Model has been quite successful in explaining the neural
phenomena. The idea of circuit equivalents and the membrane voltages
corresponding to neurons have been remarkable which is  essentially
a classical result. In view of some recent results which show that
quantum mechanics may be important at suitable length scales inside
the brain, the question which becomes quite important is to find out
a proper quantum analogue of the HH scheme which will reduce to the
well known HH model in a suitable limit. From the ideas of
neuro-manifold and the relevance of quantum mechanics at some length
scales in the ion channels, we investigate this situation in this
paper by taking into consideration the Schr\"odinger equation in an
arbitrary manifold with a metric, which is in some sense a special
case of the heat kernel equation. The next important approach we
have taken in order to bring about it's relevance in brain studies
and to make connection with HH models is to find out a plausible
circuit equivalents of it. What we do realize is that for a proper
quantum mechanical description and it's circuit implementation of
the same we need to incorporate the non commutativity inside the
circuit model. It has been realized here that the metric is a
dynamical entity governing space time and for considering equivalent
circuits it plays a very distinct role. We have used  the methods of
stochastic quantization and have constructed a specific case here
and see that HH model inductances gets renormalized in the quantum
limit.

\vspace{.5cm} {\bf Keywords :} Hodgkin-Huxely model, Quantum
Mechanics, Circuit Implementation, Ion Channel \vspace{.5cm} {\bf
PACS No.:}87.10.+e
\vspace{.5cm}

\newpage

\section{Introduction}

Nervous systems use electrical signals which propagate through ion
channels which are specialized proteins and provide a selective
conduction pathway, through which appropriate ions are escorted to
the cell's outer membrane. Also, the ion channels undergo fast
conformational changes in response to metabolic activities which
opens or closes the channels as gates. The gating is essentially
changes in voltages across the membrane and ligands. The voltage
dependent ion channels have an ability to alter ion permeability of
membranes in response to changes in transmembrane potentials. The
magnitude of current across membrane depends on the density of
channels, conductance of the open channel and how often the channel
spends in the open position or the probability. Hodgkin and Huxley
\cite{hh1,hh01} accounted for the voltage sensitivity of Na$^{+}$
and K$^{+}$conductance of the squid giant axon by postulating charge
movement between kinetically distinct states of hypothetical
activating particles. In spite of the detail electrophysiological
studies, the atomic structure of voltage gated ion channels still
remained in the dark till the discovery of Mckinnon and his
collaborators \cite{mck00,mck,mck1} which obtained a crystal
structure of a $Ca^{2+}$ gated $K^{+}$ ion channel provides a
mechanism for gating \cite{gate,gate2}.A functional study of $KvAP$
in this context led to a proposal known as the voltage sensor paddle
model.

Considering the voltage sensor capabilities of the ion channels and
generation of currents and potentials, we in this paper deal mainly
with the electrical properties of the ion channels. It is already
known that the neuron acts as an electrical device, \cite{nelec}
where a potential difference develops across the membrane due to
differences in ion concentrations between inside and outside the
cell. The participating ions are Sodium($Na^{+}$), Potassium
$(K^{+}$), Calcium ($Ca^{+}$) and Chlorine($Cl^{+}$). Nernst
equation describes equilibrium potential for a single ionic species
as

$$E = \frac{RT}{zF}ln\frac{[X^{+}]_{o}}{[X^{+}]_{i}}$$

Membrane potential due to the combined permeability of different
ionic species is given by the Goldman-Hodgkin-Katz equation
\begin{equation}
V_{m} = \frac{RT}{zF}ln\frac{K_{o}+ [p_{Na}/p_{K}]Na_{o}
        +[p_{Cl}/p_{K}]Cl_{i} }{K_{i}+ [p_{Na}/p_{K}]Na_{i} +[p_{Cl}/
         p_{K}]Cl_{o}}
\end{equation}
Total membrane current is given by the sum of individual channel
currents  $$I_{m} = I_{Na} + I_{K} + I_{Cl}$$

In this way, a membrane patch can be described by an equivalent
electrical circuit component. As we have discussed earlier,
electrical signals are changes in the membrane potential at specific
sites of the neuronal network, which are obtained by changes due the
closing and opening of ion channels. Given these things to be known,
the main objective of this paper is different. In a recent article
\cite{qmic} it has been hypothesized by some dimensional arguments
that quantum mechanics may be operative at some scale in the ion
channels. If this is the case then the whole story of voltage
sensing in ion channel gets a new paradigm shift. If we assume that
membrane voltage and currents are generated through equivalent
circuits but at length scales where quantum mechanics is assumed to
hold, then due to noncommutative effects the whole concept of
devising electrical circuits is different, but also at the same time
it should be mentioned here that at large length scales
corresponding to a large collection of ion channels in comparison to
a single or few ion channels in the previous case, we expect the
quantum effects to average out and the conventional circuit elements
for describing the mechanisms of voltage \cite{nelec1} and current
generation through the gates is valid.

In this paper, we implement a quantum circuit for the ion channels
following the lines of \cite{qcio2}. The basic task in hand is as
follows, we have developed a Schrodinger equation and a
implementation of the equivalent circuit. Now following the work
\cite{amaq} on neuromanifolds we assume that the underlying geometry
of the ion channels is not known a priori. Thereby we assume a
curved manifold and write down the nonlinear Schr\"odinger equation
(NLSE), essentially a heat kernel equation in a curved manifold
\cite{nlse00}. The next task in hand is to find out an equivalent
circuit model for that. In the last section we find out a connection
with the HH model and determine how the quantum effects may get lost
at large length scales in the mesoscopic case when we take the
limiting case of large number of ion channels.

\section{Hodgkin-Huxley Equations}
The model we would like to describe is a neuronal model at length
scales of ion channels where we believe that quantum mechanics may
be operative. But we believe that the model may also include the HH
model as a special case where coarse graining can be done, or for
example, if we include large number of channels, the collective
behaviour should be described by the HH model. For the sake of
completeness, we would like to describe the HH model in brief
\cite{hh2}.

In the HH case, the basic membrane circuit suitable for, say, a
squid giant axon with two voltage dependent channels is given by the
following construction:
 The circuit is described by a capacitor $C$,
sodium, potassium, leakage conductance $G_{Na}$, $ G_{K}$ and
$G_{L}$ respectively. The membrane potential is the voltage
difference between the outside and inside of the cell membrane and
there can be a current injected into the cell from an electrode or
other parts of the cell.

 The equations describing the phenomena is given by
\begin{equation}\label{hh1}
C \frac{dV}{dt}= I_{ext} - G_{Na}(V-E_{Na}) - G_{K}(V-E_{K}) -
G_{L}(V-E_{L})
\end{equation}

$G_{Na}$ and $G_{K}$ are the functions of membrane potentials and
time and are given by the following equations,
\begin{eqnarray}\label{hh2}
G_{Na} &=& \overline{G_{Na}}m^{3}h \quad \frac{dm}{dt} =
\frac{m_{\infty}(V) - m }{\tau_{m}(V)}\quad \frac{dh}{dt} =
\frac{h_{\infty}(V) - h }{\tau_{h}(V)}\\ \nonumber G_{K} &=&
\overline{G_{K}}n^{4} \quad \frac{dn}{dt} =
\frac{n_{\infty}(V) - n }{\tau_{n}(V)}\\
\end{eqnarray}
Here $m^{3}h,  n^{4}$ can be interpreted as the opening probability
of a channel. The $Na$ channel has two set of gates i.e., activation
gates represented by $m$ and inactivation gates represented by $h$.
The activation gates open and the inactivation gates close when the
membrane depolarizes. The $K$ channel has only single activation
variable which is a $4$ parameter system.

So we see that the state vector variables of the HH model are
$V,m,h,n$. The equations[2,3] and [4] can be written in a compact
matrix notation as
\begin{equation}\label{hh3}
\dot{\overrightarrow{X}} = \overrightarrow{F}(\overrightarrow{X})
\end{equation}
where $\overrightarrow{X} = [V,m,h,n]^{T}$. Equation [\ref{hh3}] is
a nonlinear equation and mainly numerical methods are employed in
solving such equations. We will not go into details of those
analysis as here we are interested to carry out the analysis in
terms of the relevance of quantum mechanics in ion channels and
develop a framework for that and then to see if there exist any
limit for which it will reduce to the HH model. More interestingly,
the observation that understanding ion channel dynamics is
stochastic in nature has prompted us to look at the relevance or
analog of stochasticity in the quantum case. A similarity of the HH
model with the cellular automation has been observed,
\cite{fox,fox1,defil1} which in the limit of large ion channel
density gives rise to a Langevin description. Using the Stratonovich
description, the HH model is rewritten in the Langevin form as

\begin{equation}
\frac{d}{dt}x_{i}= A_{i}(x)+ B_{ij}\eta_{j}(t)
\end{equation}
where $i,j = 1 \cdots n$ for the $n$ channels and $A_{i}, B_{ij}$
are related with the moments of the underlying transition
probability.

It is striking that HH formulation yields into a noisy model in the
large ion channel number limit. This observation has become very
crucial in our proposal of the general formulation of the HH
formalism in the quantum case.

\section{\bf Nonlinear Shr\"odinger Equation (NLSE) and the \\
relation of Stochastic geometry with \\
Neural  Modeling }\label{secstoch}

Very Recently the NLSE has been solved with an artificial neural
network scheme. This analysis gives us an insight and confidence
that maybe the NLSE will play an important role in analyzing
realistic neuronal modeling. Here we discuss in brief about the
solution of NLSE on a network for the sake of the completeness of
our proposed argument.

The time dependent propagation of light pulse inside a single mode
nonlinear optical fiber is given by the solution of
\begin{equation}
i(\frac{\partial \Psi}{\partial z}) -\alpha (\frac{\partial^{2}
\Psi}{\partial z ^{2}}) -
 \beta {\Vert{\Psi}\Vert}^{2}\Psi = 0
\end{equation}
where $\Psi$ is the field amplitude, $z$ and $t$ are the optical and
time axis respectively, $\alpha, \beta$ are the dispersive and
waveguide coefficients respectively. The competition between pulse
dispersion and focussing gives rise to the formation of solitons for
a particular input. With suitable boundary conditions a stable
soliton is obtained. It has been observed that the solution consists
of a 3 layer architecture with 42 hidden nodes \cite{nn1}. Now to
speak of the implications of this result it has been also observed
that the knowledge of the upper bound on the field amplitude
provides a stopping criterion on the training of the neural network
(NN).

It may be pertinent to ask at this stage that what use is of the
above scheme to our proposed model. What we believe is that
applicability of NLSE on NN gives us a clue that may be the
Schr\"odinger equation is applicable at some length scales in
neuronal architecture with an unknown, a priori geometry and the
basic objective is to find out the appropriate dynamics for that.

We have already emphasized that the neuronal architecture has a form
of geometry with some probabilistic structure on it giving rise to a
probabilistic manifold \cite{prob}. So the main point of the
analysis depends on the identification of a stochastic
interpretation to quantum mechanics. The essential ingredient is
following. We claim that the
 operator $A = b_{\nu}(x)\partial_{\nu} + (\frac{\hbar}{2\pi i}) {\nabla} $ is the
  infinitesimal  generator of the stochastic process defined by the Langevin equation
\begin{equation}
dx_{\mu}(t) = b_{\mu}(x,t) dt + dW_{\mu}(t)
\end{equation}
The importance of this identification is that classical probability
theory gets related with quantum mechanics. Now, the next question
what we can ask is that we are trying to define the SE in a curved
probabilistic manifold. So, apart from a stochastic approach to
quantum mechanics ,we need something more, i.e., to randomize the
metric. Let us assume a Lagrangian, given by

\begin{equation}
L = \frac{m}{2}g_{\mu\nu}(x) {\dot{x}}_{\mu}{\dot{x}}_{\nu} -V(x)
\end{equation}

 Variation of this Lagrangian gives us the equation of motion in the form of geodesics.
  If we vary the trajectories and define a stochastic process in terms of the variations
  with gaussian spread and compare this distribution with the Feynman path integral, we end
  up with the Riccati equation which is the stochastic analogue of the Schr\"odinger equation
   on a curved manifold.
 \begin{equation}
 \label{stoch1}
\frac{\hbar}{2}{\nabla}_{\mu} x^{\mu} + \frac{m}{2}x^{\mu}x_{\mu} =
V + \frac{{\hbar}^2R}{6m}
\end{equation}

So we see that stochasticity \cite{stoch00} involves a generation
of an effective potential of motion. We will see now that how this
may be handled in analyzing the quantum circuits.

\section{Circuit Implementation Of Schr\"odinger \\
Equation }

It is indeed true that in understanding neural mechanisms, we need
nonlinear and dissipative analysis. Now as has been argued in
\cite{qmic3} over the years and until recently, if we think of the
relevance of quantum mechanics for neuronal dynamics at suitable
length scales we should try our hands on quantum dissipative
systems. Quantum mechanical systems are inherently open systems. In
an open system ,for example, in an atom in a cavity, a process such
as spontaneous emission is sometimes viewed as dissipative but if
the number of modes is reduced, the process becomes reversible. In
comparison, the resistance to electric current flow is reversible,
which is typical of closed system. But if we think of quantum
circuits the situation is drastically different.

In this context, we cite an particular example: A current driven RC
circuit which is identical to a free particle driven by an external
force. In absence of the resistor, the system is well described by
the charge operator $Q$ and operator $\phi$ which satisfy $ [\phi,
Q] = i\hbar$. We would like to mention here that a circuit theory
\cite{ckt00,ckt} that can describe quantum transport, is
particularly important and has potential applications in
nanotechnology, molecular devices and beam epitaxy etc \cite{nano}.

As we have already seen that the electric charges of ions are in
fact responsible for the membrane potential and action potential.
Generation of the potential therefore gives rise to the possibility
of modeling the ion channels through electric circuits, which
generate the required potentials. The scheme is devised through a
quantum analogue of the corresponding electrical circuit. So our
objective is pretty clear. Some very recent results at ionic scales
regarding the relevance of Quantum Mechanics (QM) \cite{qmic4}, we try to build
some viable neuronal models and corresponding electric circuits \cite{ckt3}. But
as QM governs the dominant dynamics we have to develop quantum
circuits.

In this context, we would like to mention very important work
\cite{qcio2} which we earlier mentioned, related to the circuit
equivalent of Schrodinger's equation. The circuits were originally
designed for completely different purposes and had no connection
with brain activity. It is a way to measure the eigenvalues,
eigenvectors and statistical means of various operators, belonging
to the system which are being modeled by electrical means. We
briefly discus below the scheme for handling those things.

Let the wave equation be divided by $i\omega_{c}$ where $\omega_{c}
= \sqrt{\omega} = (E/\hbar)^{\frac{1}{2}}$ and multiplied by
$\triangle x$ we get
\begin{equation}
-\frac{1}{\omega_{c}}\frac{\hbar^{2}}{2m}\frac{1}{\triangle
x}\frac{\partial^{2}\psi}{\partial x^{2}}\triangle x^{2} + \Big(
\frac{V\triangle x}{i\omega_{c}} + \hbar i \omega_{c}\Big) \psi = 0
\end{equation}

We should like to note some salient points here.
 \begin{itemize}
\item The Kinetic energy operator $T$ is represented by a set of
inductors in series whose inductance is given by $ L_{1} =
\frac{2m}{\hbar ^2}\triangle x$.

\item The potential energy operator
V is represented by a set of unequal coils in parallel whose
inductance is $L_{2}= 1/V\triangle x$.

\item The total energy
operator $- E$ is represented by a set of equal capacitors whose
capacitance is $\hbar\triangle x$.

\item The operand $\psi$ is
represented by voltages and the result of the operation $\alpha
\psi$ where $ \alpha$ is any operator, is represented by currents.
 \end {itemize}

This model can also be extended to nonorthogonal coordinates and in
general on arbitrary manifolds. Utilizing the circuits, tests were
carried out on an ac network analyzer. The results are worth
mentioning. The tests were done in 1-dimensional circuits. or
example measurements were made for a particular case of the
rectangular potential well and analyzed which had good agreement
with the experimental results \cite{exp}.

For the sake of completeness, we should like to mention here that
from the preceding model, we can develop a prescription for
developing a electric circuit equivalent for the Schr\"odinger
equation (SE). The SE has some analogies with the heat conduction
equation

\begin{equation}
\frac{d\psi}{dt} = \frac{1}{\hbar}\Big(\frac{\hbar^{2}}{2m}
\nabla^{2} + V \Big)\psi
\end{equation}

Now we make a prescription for the electric circuit as equivalent to
the Schr\"odinger equation as

\begin{eqnarray}
V\Leftrightarrow \psi \quad \frac{1}{r}\Leftrightarrow V\triangle
Vol \quad \frac{1}{R x}\Leftrightarrow \frac{\hbar^{2}}{2m}
\triangle Vol/(\triangle x^{2}) \quad C\Leftrightarrow i\hbar
\triangle Vol
\end{eqnarray}

The construction relies here on having $N$ imaginary capacitance and
one of the consequence is we will get solutions of the form

$$ \psi\sim \exp(ikx) \exp(i\omega t)$$.
 Physically this means that we don't
have exponential decay with time into thermal equilibrium but we get
everlasting solutions which conserve $|\psi|^{2}$. This may sound
somewhat unrealistic and there are some ways of circumventing this
problem.

\section {Quantum Mechanics in Ion Channels and \\
the Formulation}

Motivated by the quantum mechanical considerations and the circuit
equivalence of Schr\"odinger equation we will now try to formulate
an equivalent circuit for the membrane potential in the ionic
channels. We will consider a single ion channel and consider the
circuit implementation for it. But there are some subtleties
regarding this. Tensor network theory \cite{tnet} may be realized in
the brain and there is possibility  of a non trivial geometrical
structure inside brain as mentioned by Amari \cite{ama12}, Roy and
Kafatos \cite(qmic5). Pribram's \cite{prib} work also points towards
this direction. This implies that there is an underlying geometrical
structure inside brain and it may be important at the ionic scales.
So we try to develop a formalism for describing that. We start with
some simplifying propositions:
\begin{itemize}

\item  There is an underlying geometry inside brain which is
responsible for neuronal activities and the geometry can be
described by a metric.
\item  Quantum mechanics is applicable at the length scales of ionic
channels and the phenomena can be described by Schr\"odinger
equation
\item  The phenomena at those length scales is stochastic.
\end{itemize}

With these propositions we can now think of a formalism for
various set of events inside the brain. It should be mentioned
here that ultimately we would like to connect our formalism to
the HH formalism, which has been successful in describing the
membrane gating and dynamical phenomena involving channels
\cite{ic}.

So we start by writing a SE equation on a curved manifold. The
equation can be identified with a Heat Kernel equation.
\begin{equation}
i\hbar\frac{\partial \psi}{ \partial t} =
\frac{1}{\sqrt{g}}\partial_{\mu}(\sqrt{g}\partial_{\mu}\psi) + V\psi
\end{equation}
The equation above, for arbitrary metric, is in general, nonlinear
and in accordance with our third proposition we claim that the
processes are stochastic in nature and the analysis of section
\ref{secstoch} tells us that the dynamics will be governed by
Riccati like equation [\ref{stoch1}] with a correction in the
potential term as

\begin{equation}
i\hbar\frac{\partial \psi}{ \partial t} = \frac{\hbar^{2}}{2m}
\nabla^{2}\psi + (V + \frac{\chi (q)
\hbar^{2}\sqrt{N}\vartheta}{6})\psi
\end{equation}

$\vartheta$ is the curvature scalar, associated with the metric,
which gets incorporated into the potential term. It is important to
note that the correction term differs from that of equation
[\ref{stoch1}] in a coordinate dependent factor. We will show that
this factor is crucial in developing a quantum equivalent of a
circuit. The mass is absorbed in the coordinate dependent term. The $\chi$ in some sense acts as a space
modulation factor which governs both the noncommutative aspects and the fluctuations from the metric.It
should be mentioned here that this equation describes the dynamics
for $N$ channels and we have made a conjecture by including the
number of channels, with a hope to get the classical picture of the
HH equation in global limit.

The Schr\"odinger equation along with the correction of the quantum
term is a good starting point in our case to develop a circuit
equivalent. Actually the problem in this case to extend the previous
construction for the equivalence of SE to electrical circuit is
dictated by the presence of the metric. First of all, the metric is
a dynamical variable which governs the behavior of space time. So,
we cannot just implement it as some electrical component, since such
a component should have the ability to shape the global structure of
the full circuit. At the moment, we do not know of any such
component. We have seen that in periodically driven circuits, we can
scale the capacitance or say the inductance as $ L \sim \gamma g(t)
$ which may capture, in some sense, the global dynamics, but it will
not catch the full glimpse of the dynamical behavior. The curvature
gets into the potential term and thereby fluctuations in the metric
will induce different potentials and hence only periodic variations
may not do \cite{sr11,sr22}. It is a widely held that fundamental
processes of nature may be explained by probabilistic metric and the
probabilistic features can be modeled into uncertainties or
fluctuations from a physical point of view. If we introduce the
fluctuations in the metric as
$$ g_{ij}(x, h) = g_{ij}(x) + \alpha_{ij}(h) $$
 the fluctuation of the
metric generates a random potential $ V$, a random coefficient $S$
which depends on the fluctuations. In the quantum case we will do
indeed get dissipation which depends on the fluctuation. The
Schr\"odinger equation turns to be
\begin{equation}
\frac{\hbar^{2}}{2} \frac{\partial^2 \phi}{\partial r^{2}} + V\phi =
 S\frac{\partial \phi}{\partial t}
\end{equation}

To make connections with brain activities and neuronal circuits we
try to develop circuits corresponding to quantum mechanics, the
circuit will do contain some flavor of the noncommutative aspects.

We know that brain phenomena is considered as dissipative. In such
kind of theories, one considers such one partcle dissipation in
quantum theory. So, we try to extend that formalism in our case with
the corrected potential along with a source term. Then we consider
the following Hamiltonian as:
\begin{equation}
H = -e^{-Rt/L}\frac{\hbar^{2}}{2e^{2}L}p^{2} +
e^{Rt/L}(\frac{1}{2C}q^{2} + \varepsilon q +
\frac{\chi(q)\vartheta \hbar^{2}\sqrt{N}}{6})
\end{equation}
Here $q$ is state variable, $p$ the conjugate which goes uplifted to
the charge operator when we deal with quantum mechanics (QM). Using
the Heisenberg equation of motion, the equation for the state
variable is given by
\begin{equation}
L\frac{d^{2}q}{dt^{2}} = \frac{1}{2e^{2}}(\{\frac{1}{2C}q^{2}
+\varepsilon q + \frac{\chi(q) \vartheta \hbar^{2}\sqrt{N}}{6}\},
[p^{2}, q]) - R\dot{q}
\end{equation}
So, evaluation of the simple commutator gives us the equation for
the corresponding quantum circuit as
\begin{equation} L\ddot{q} +
R\dot{q} + \frac{q}{C} = \varepsilon +
\frac{\alpha\hbar^{2}\sqrt{N}}{6}\vartheta \dot{q} \int {dt
\chi'(q)}
\end{equation}
The last term  in the above equation is most striking. It shows that
that the above equation clearly shows that the inductance gets
corrected, by a quantum term. In this way, ultimately, at the level
of circuit equivalence, we will be getting a renormalized
inductance. It gives the equation a status of an integral equation
and would be interesting to find out the conditions under which it
will reduce to a differential equation. In that case, the
capacitance gets renormalized.

It is very important so as to make some measurements to find out
these extra factors. There is an extra parameter in the theory which
needs to be fine tuned to get the desired effects. The above
equation can also be transformed into a Langevin like form and get a
measure for the Probability functional, which is very non trivial
due to the presence of the curvature term and may hint at some
statistical manifold like character. This is not quite surprising as
we mentioned at the beginning of the section that it may arise due
to the intrinsic stochasticity of the neuronal activities. The above
observations have some interesting consequences with respect to
Nonlinear Schrodinger Equation. Some of the results is worth
mentioning . If we had included in the Schr\"odinger equation a
damping factor in the form of a bounded negative operator and a
quasi periodic force, the solutions turn out to be even and
periodic. The analysis in such case, give rise to the existence of
invariant manifolds in the phase space of the equation. The
infinitely many eigenvalues in the integrable limit turn into
complex eigenvalues with negative real parts. The manifolds exhibit
a dynamical behavior and the geometry resembles those of certain
homolinic orbits in finite dimensional Ordinary Differential
Equation (ODE \cite{slink}).

\section{\bf Discussions}

The striking aspect of our result is that in the most general case,
for scales in which QM is applicable, we have found out a
generalized HH equation with the conductances $ G_{A}$ being
corrected with the renormalized in equation [\ref{hh1}] by

$$ G_{A}+ ({\frac{\alpha\hbar^{2}\sqrt{N}}{6}\vartheta \dot{q}\int {dt
\chi'(q)}})^{-1}$$

It is necessary to study following two issues. Firstly, to see under
what limit does this modified equation i.e., generalized HH equation
turns to ordinary equation with no renormalization. Then, one needs
to do the experiments to see whether the conductances indeed do get
corrected. If it is so then  we could measure such term for single
ion channels. We also believe that one of the mechanisms by which we
may get ordinary HH theory with no renormalization (i.e quantum
mechanics is unimportant) is when there are many channels and
quantum mechanics is getting subdued in the large $N$ limit. Anyway,
there is a subtle point here. In confirmation of the relevant
observation for the stochasticity of HH equation in the Langevin
description, we see that in the classical limit, we may get
stochasticity for a critical large value of the number of channels.
It is really important to design experiments to measure critical
parameters as appeared in the above equation. Such experiments will
be very conclusive for the correctness of the model and also give a
direct evidence for the applicability of QM in ion channels. It is
also to measure the effective conductance. At this stage, we still
do not know how we can model such mechanisms, but experimental
results on single ion channel may surely shed some light in
understanding these aspects \cite{exp00,exp001}.

\end{document}